\documentclass[prb,aps,twocolumn,showpacs,amsmath,amssymb]{revtex4}
\usepackage{graphicx}
\usepackage{dcolumn}

\begin{document}

\title{Two-Stage Kondo Effect and Kondo Box Level Spectroscopy in a Carbon Nanotube}

\author{Yu. Bomze, I. Borzenets, H. Mebrahtu, A. Makarovski, H. U. Baranger,  and G. Finkelstein}

\affiliation{Department of Physics, Duke University, Durham, NC 27708}

\date{August 7, 2010}

\begin{abstract}
The concept of the ``Kondo box'' describes a single spin, antiferromagnetically coupled to a quantum dot with a finite level spacing. Here, a Kondo box is formed in a carbon nanotube interacting with a localized electron. We investigate the spins of its first few eigenstates and compare them to a recent theory. In an `open' Kondo-box, strongly coupled to the leads, we observe a non-monotonic temperature dependence of the nanotube conductance, which results from a competition between the Kondo-box singlet and the `conventional' Kondo state that couples the nanotube to the leads.
\end{abstract}

\maketitle

One of the exciting concepts that has emerged in modern condensed matter physics is that of competing many-body ground states: the existence of states with different symmetries leads, for instance, to distinct phases separated by quantum phase transitions \cite{Sachdev}. It was later appreciated that similar physics also occurs in an individual quantum impurity system \cite{Vojta_review}, where distinct many-body states may for example possess different spin. Experimentally, this physics is best studied in nanostructures, in which one can tune the system parameters and create complex set-ups with competing many-body states \cite{Chang_Chen_2009}.

The quantum impurity system that we study is a carbon nanotube quantum dot exchange-coupled to a localized electron [schematic in Figure 1(b)], thereby forming a ``Kondo box'' \cite{Thimm99,early_box,Ribhu,Pereira2008}. We focus on the case of an odd number of electrons in the nanotube and identify the spin of the ground state and several excited states. For a Kondo box strongly coupled to the leads, the nanotube spin can either be Kondo-screened by the leads or form a singlet state with the localized electron. We find that the competition between these distinct ground states results in a peculiar non-monotonic dependence of the nanotube conductance on temperature, where the initial rise in conductance at intermediate temperatures changes into a sharp drop at the lowest temperatures. At these lowest temperatures, a single channel in the leads screens both the localized spin and the spin in the nanotube quantum dot \cite{Vojta}.


\begin{figure}[t]
\includegraphics[width=.8 \columnwidth]{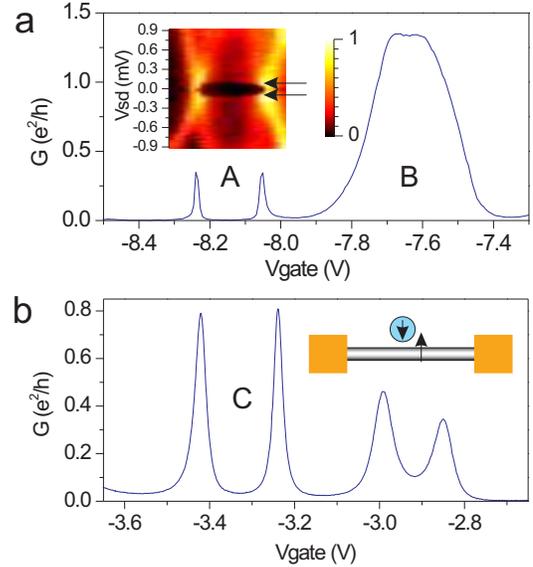}
\caption{\label{fig:2valleys}
(color online) 
(a), (b) Zero-bias differential conductance as a function of $V_{\rm gate}$ showing 4 single-electron conductance peaks in each panel. We focus our attention on three ``odd-electron'' valleys  `A', `B' and `C'. The two peaks bordering valley B are almost completely merged due to the `conventional' nanotube-leads Kondo effect \cite{Kondo1998}.
Inset in (a): differential conductance map of valley A measured \emph{vs.} $V_{\rm SD}$ and $V_{\rm gate}$ (shared horizontal axis with the main panel). The arrows indicate the onset of the inelastic cotunneling processes. The colored bar shows the 0 to 1 $e^2/h$ scale; the same type of colormap is used throughout the text. 
Inset in (b): schematic of the nanotube with an odd number of electrons and an extra electron localized nearby.
}
\end{figure}


The nanotubes are grown on a Si/SiO$_2$ substrate by chemical vapor deposition using CO as a feedstock gas \cite{Zheng2002}. This method usually produces nanotubes with diameters of about 2 nm. We present results measured on a single semiconducting nanotube with two metal contacts separated by a distance of 400 nm. Similar results have been observed in other samples. The measurements were performed at the base temperature of 25 mK (except for the temperature dependence in Fig.\ 3).

Figures 1(a) and (b) show the nanotube zero-bias differential conductance as a function of gate voltage ($V_{\rm gate}$). Both graphs correspond to adding 4 successive electrons to the nanotube, as seen by 4 single-electron conductance peaks. (The feature `B' in Figure 1(a) is formed by two peaks merged by the `conventional' Kondo effect \cite{Kondo1998}; they can be separated by a finite source-drain bias.) We will focus most of our attention on valleys `A' and `C', each of which is narrow and flanked by two peaks of comparable height and width. These signs point to valleys with an ``odd-electron'' occupation: the two peaks correspond to adding spin-up/down electrons to the same quantized level; the valleys are narrow since their widths are proportional to just the charging energy, while the neighboring even-electron valleys have an additional contribution from the quantization energy \cite{QDreview}. The orbital (shell) degeneracy \cite{4peaks} is broken in this sample. From differential conductance spectroscopy we have identified the following energy scales: the charging energy is $\sim $3-4 meV, shell spacing is $\sim 2$ meV, and orbital spacing within a shell is $\sim$200-300 $\mu$eV (valley A) or $\sim$1 meV (valley C).


Odd-electron valleys usually demonstrate a zero-bias resonance (large conductance, as observed in valley B) associated with the Kondo state formed between the quantum dot and its leads (referred to here as the ``nanotube-leads Kondo state'') \cite{Kondo1998,Pustilnik&Glazman2004}. Instead of such a Kondo resonance, the map of valley A [inset of Fig.\ 1(a)] shows an unexpected {\it suppression} of conductance, which extends up to $V_{\rm SD} \approx \pm 0.1$ meV, at which point an inelastic cotunneling threshold is visible. A similar gap is also observed in valley C and several other, but not all, odd-electron valleys. The cotunneling gap indicates the presence of a low-energy excited state. 

Figure 2(a) shows the evolution of this cotunneling gap in magnetic field perpendicular to the nanotube, $B_{\perp}$, for the middle of valley A ($V_{\rm gate}$ fixed at $-8.148\,V$). Upon application of $B_{\perp}$, the gap becomes narrower, showing that the energy difference between the ground and the excited states decreases. It closes at $B_c \approx 1$ T, and reopens at higher $B_{\perp}$. Valley C shows a similar behavior with $B_c \approx 0.3$ T. These observations indicate that the ground state and the excited state cross at $B_c$; thus, the excited state has a larger spin than the ground state. 

\begin{figure}[t]
\includegraphics[width=1 \columnwidth]{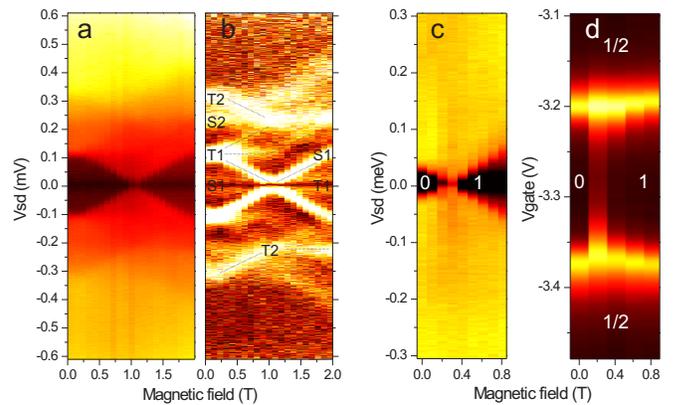}
\caption{\label{fig:cotunneling}
(color online) 
(a) Differential conductance map measured \emph{vs.} $V_{\rm SD}$ and magnetic field perpendicular to the nanotube, $B_{\perp}$. $V_{\rm gate}$ is fixed at $-8.148$ V corresponding to the middle of valley A in Figure 1a. Colormap scale: 0 to 0.07 $e^2/h$. The features observed at finite $V_{\rm SD}$ are inelastic cotunneling thresholds. The most prominent threshold at 0.1 meV corresponds to the singlet-triplet gap, which closes at $B \approx 1$ T. 
(b) Numerical derivative of the conductance map (absolute value, arbitrary units). The three components of T1 are visible, as well as S2 and T2. Above $B \approx 1$T the lines change slope because the ground state changes from S1 to the lowest component of T1. Zero-bias suppression seen on the scale of tens of $\mu$V is due to environmental blockade in the resistive leads to the nanotube \cite{dissipation}. This feature appears at smaller energy and independently of any nanotube-specific features discussed here. (c) Differential conductance map measured \emph{vs.} $V_{\rm SD}$ and $B_{\perp}$ in the middle of the valley C. The singlet-triplet crossing occurs at $\sim 0.3$ T. Colormap scale: 0.05 to 0.15 $e^2/h$. (d) Zero-bias differential conductance map measured \emph{vs.} $V_{\rm gate}$ and $B_{\perp}$ in valley C. Colormap scale: 0 to 0.7 $e^2/h$. 
}
\end{figure}

Based on these observations, the natural candidates for the ground and excited states are a spin singlet (S1) and a spin triplet (T1) formed by the odd number of electrons in the nanotube and an extra, antiferromagnetically coupled electron localized close to the nanotube. Indeed, behavior of a cotunneling threshold similar to that in Fig.\ 2 has been observed for the singlet-triplet gap in quantum dots with an even-electron occupation \cite{Zumbuhl}. To better observe the excited states, we take a numerical derivative of the data of Fig.\ 2(a). The  resulting Fig.\ 2(b) immediately confirms our assignment of the first excited state (T1) as a triplet: it splits three ways in $B_{\perp}$, where the three components go up, down, or stay flat. We can also observe the two higher excited states: another singlet (S2) and triplet (T2). This sequence of excited states is consistent with the theoretical predictions of Ref.\,\onlinecite{Ribhu}, in which it is shown that the ground state of the Kondo box is a singlet (S1) for any value of the antiferromagnetic exchange coupling. In both the limit of a very strong or very weak coupling, it is followed by T1 and S2, so the authors surmised that the same sequence of states persists for any strength of coupling. In our case, the gaps between S1, T1, and S2 are similar, so the exchange coupling is comparable to the orbital splitting. 

The unexpected formation of a cotunneling gap instead of a Kondo resonance has been reported earlier in nanotube samples \cite{BabicGap,Nygard}. Some of these gaps have been attributed to the influence of magnetic quasiparticles used as a catalyst for the nanotube growth. At least in one of the reported cases \cite{Nygard}, the cotunneling gap \emph{grows} rather than closes in magnetic field. They found that the gap value is asymmetric with respect to applied magnetic field, possibly due to the frozen magnetization of the catalyst particle. We can exclude such a scenario here because the dependence of the gap on magnetic field is symmetric around zero without hysteresis. Furthermore, the $g$-factor of the extra electron is very close to $2$ in both valleys A and C [Figs.\ 2(a) and (c)], suggesting an electron localized in SiO$_2$ or a light adatom, readily available in nanotubes exposed to the ambient conditions. 

The exchange interaction between the topmost electron in the nanotube and a covalently attached adatom should be greatly reduced compared to the eV atomic-scale exchange. Indeed, the nanotube electron is spread over $10^4$ to $10^5$ carbon atoms, bringing the strength of the exchange coupling to the $\mu$eV range that we observe in valleys A and C. Alternatively, the electron may be residing in the substrate within a sub-nanometer proximity to the nanotube. It is known that large gate voltages applied to the nanotube can inject electrons from the nanotube into SiO$_2$ \cite{Fuhrer}.

Experimentally, we have observed a variety of singlet-triplet gap values; they do not seem to obey a clear pattern as a function of $V_{\rm gate}$. The random overlap of the nanotube electron with the localized state and the successive filling of different localized states should both cause irregular variations of the exchange interaction with $V_{\rm gate}$. Furthermore, the size of the singlet-triplet gap is renormalized from the bare exchange interaction \cite{Ribhu}. 

The presence of the extra electron shows up not only in the cotunneling spectra, but also in the zero-bias conductance. In Figure 2(d), we study in a magnetic field the positions of the conductance peaks that border valley C. The peaks first come closer together before spreading apart. The change in the slope occurs when the spin in valley C changes from 0 to 1, while the spin in the neighboring valleys stays $1/2$, (due to the localized electron). Indeed, the transition field coincides with the closing of the singlet-triplet gap seen in the cotunneling spectrum [Fig.\  2(c)]. Overall, this behavior is similar to the singlet-triplet transition observed in quantum dots at even filling \cite{QDreview}, but with the very important difference that here the number of electrons in the nanotube itself is \emph{odd}. 

We are now ready to present the main result of this paper. First, note that the conductance in valley B [Fig.\ 1(a)] is enhanced due to the Kondo effect \cite{Kondo1998,Pustilnik&Glazman2004}, where the nanotube spin is screened to form a Kondo singlet with the leads. In valley A, the nanotube electrons instead form a singlet (S1) with the localized spin. Evidently, these two singlets compete---an example of the competing many-body grounds states mentioned in the introduction---and depending on the relative strength of the two interactions, the nanotube conductance is either suppressed by formation of S1 (valley A) or enhanced by the nanotube-leads Kondo effect (valley B). 

We ask, then, if the ground state of our Kondo-box system coupled to the leads must always be a spin singlet as in valley A. The enhanced conductance in valley B hints at a different possibility. Namely, the conventional counting of the spin degrees of freedom (two spin 1/2) and the available screening channels (one linear combination of the two leads) \cite{Vojta_review,Chang_Chen_2009,Pustilnik&Glazman2004} suggests that once the nanotube spin is Kondo-screened by the leads, the localized spin should decouple from the system. The resulting doublet ground state would be separated from the Kondo-box singlet by a quantum phase transition \cite{Chang_Chen_2009}. However, it has been demonstrated theoretically that the ground state of this system must always be a singlet \cite{Vojta}. It is further explained that at low enough temperature the spin of the localized electron should be Kondo-screened by the heavy quasiparticles of the nanotube-leads Kondo system, resulting in eventual \emph{suppression} of the conductance at very low temperature \cite{Vojta,2stage}. This behavior is observed here for the first time.

\begin{figure}[t]
\includegraphics[width=.95 \columnwidth]{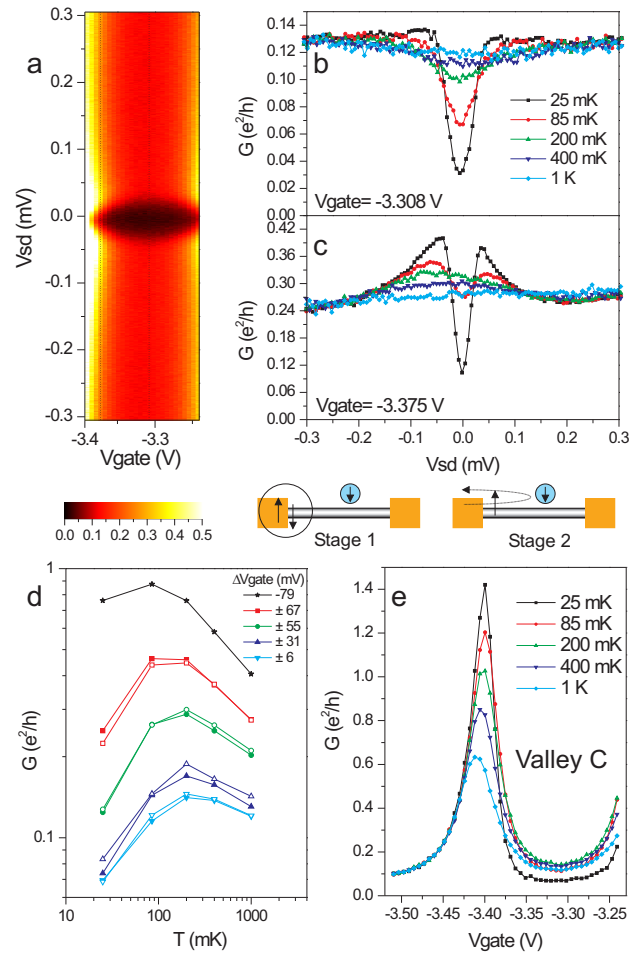}
\caption{\label{fig:2stage}
(color online) (a) Differential conductance map measured \emph{vs.} $V_{\rm gate}$ and $V_{\rm SD}$ in the Coulomb blockade valley C at the base temperature. 
(b) and (c) Differential conductance \emph{vs.} $V_{\rm SD}$ at several temperatures for $V_{\rm gate}= -3.308\,$V and $-3.375\,$V, respectively. [These $V_{\rm gate}$ are indicated in (a) by dotted lines.] Schematics: the two stages of screening the localized and the nanotube spins. 
(d) Temperature dependence of the zero-bias conductance measured at several gate voltages. $\Delta V_{\rm gate}$ indicates the gate voltage as measured from the center of the valley ($-3.308\,$V). Solid and open symbols correspond to negative and positive $\Delta V_{\rm gate}$, respectively; as expected, members of a pair show very similar behavior. (e) Zero-bias conductance \emph{vs.} $V_{\rm gate}$ at different temperatures \cite{note}. 
}
\end{figure}


The temperature range for the second stage may be experimentally unattainable in valley B; instead, we test the predicted behavior in valley C of Fig.\ 1(b). Its conductance map \emph{vs.} $V_{\rm gate}$ and $V_{\rm SD}$ is shown in Figure 3(a). We compare the temperature evolution of $G( V_{\rm SD})$ at two values of $V_{\rm gate}$: in the middle of valley C [panel (b)] and closer to its edge [panel (c)]. The nanotube-leads exchange coupling $J_1$ strongly depends on $V_{\rm gate}$ within the same Coulomb valley: it is smallest in the middle of the valley and becomes large at the edges \cite{Pustilnik&Glazman2004}. At the same time, the exchange between the nanotube and the localized state $J_2$ should not significantly change on this scale. Therefore, by comparing the spectra at two gate voltages, we vary the relative strength $J_1 / J_2$. In Fig.\ 3(b)---small $J_1$---the nanotube spin evidently couples with the localized state to form a singlet, resulting in formation of the singlet-triplet gap similar to that in valley A. The situation is different in Fig.\ 3(c)---large $J_1$---where at intermediate temperatures (200 mK) a wide resonance is formed. As the temperature is lowered, the resonance grows, except for small $V_{\rm SD}$ where a sharp dip develops at the lowest temperatures.


We attribute the initial rise of the zero-bias conductance in Figure 3(c) to the Kondo-screening of the nanotube spin by the leads; the relatively wide conductance resonance ($\sim\! 100 \mu$V) reflects the characteristic Kondo scale. Eventually, the leftover spin of the localized electron is also screened, resulting in a narrow zero-bias suppression at the lowest temperatures. Phenomenologically, a similar two-stage behavior has been observed in quantum dots in certain even-electron valleys \cite{2stage_expt}, and in nanoparticles antiferromagnetically coupled to magnetic impurities in the leads \cite{Heersche2006}. Conceptually, however, there is a critical difference: in our case, both stages of Kondo screening are performed by the \textit{same} screening channel (a linear combination of the leads)\cite{Vojta}, while in other cases two channels are required, one for each stage \cite{Pustilnik&Glazman2004}. 

Strikingly, the width of the dip in the cotunneling spectrum measured in the two-stage case is significantly narrower than that observed in the middle of the Coulomb valley [compare Fig.~3(c) to Fig.~3(b) and also note the shape of the cotunneling feature \emph{vs.} $V_{\rm gate}$ in Fig.~3(a)]. 
The value of this gap reflects the second stage Kondo temperature, which according to Refs.\,\onlinecite{2stage} should decrease with increasing $J_1$.
This behavior is evident from Fig.~3(d), which plots the zero-bias conductance as a function of temperature for several values of $\Delta V_{\rm gate}$ (gate voltage measured from the center of the valley). For larger values of $J_1$ (upper curves, larger $\Delta V_{\rm gate}$) the maximum shifts toward lower temperature, indicating the reduced temperature of the second Kondo stage. Finally, Figure 3(e) shows the complementary view of the zero-bias conductance as a function of $V_{\rm gate}$ for several temperatures. The right-hand side of the single-electron peak at $V_{\rm gate} = -3.4\,$V, which faces valley C, grows at low temperature indicating the development of the first stage---the nanotube-leads Kondo effect. Elsewhere in valley C, the conductance shows a two-stage Kondo behavior---the initial increase of conductance at high temperature is followed by eventual decrease at the lowest temperature \cite{note}. Interestingly, at the base temperature $G( V_{\rm gate})$ is rather flat in valley C. 

In conclusion, we realized a Kondo box consisting of a large carbon nanotube quantum dot coupled to a localized electron. For an odd number of electrons in the nanotube, we studied the cotunneling features and determined the spin of the ground and several excited states, which were found to agree with the theoretical predictions \cite{Ribhu}. For the Kondo box strongly coupled to the leads, we discussed the competition between the Kondo-box singlet and screening of the nanotube spin by the leads. The competition leads a two-stage Kondo effect in a situation in which there is a single screening channel: the initial enhancement of conductance gives way to eventual suppression at the lowest temperatures when the heavy quasiparticles finally screen the localized spin. 

We thank Ian Affleck, Albert Chang, Eran Sela, and Denis Ullmo for valuable discussions. This work was supported by DOE DE-SC0002765.


\vspace*{-0.1in}
\begin{thebibliography}{99}

\bibitem{Sachdev} S. Sachdev, Quantum Phase Transitions (Cambridge University Press, 2001).

\bibitem{Vojta_review} M. Vojta, Philosophical Magazine {\bf 86}, 1807 (2006).

\bibitem{Chang_Chen_2009} A.M. Chang and J.C. Chen, Rep. Prog. Phys. {\bf 72}, 096501 (2009).



\bibitem{Thimm99} W.B. Thimm, J. Kroha, and J. von Delft, Phys. Rev. Lett. {\bf 82}, 2143 (1999).
 

\bibitem{early_box} P. Simon and I. Affleck, Phys. Rev. Lett. {\bf 89}, 206602 (2002); P.S. Cornaglia and C.A. Balseiro, Phys. Rev. Lett. {\bf 90}, 216801 (2003).

\bibitem{Ribhu} R.K. Kaul, G. Zarand, S. Chandrasekharan, D. Ullmo, and H.U. Baranger, Phys. Rev. Lett. {\bf 96}, 176802 (2006); Phys. Rev. B {\bf 80}, 035318 (2009).

\bibitem{Pereira2008} R.G. Pereira, N. Laflorencie, I. Affleck, and B.I. Halperin, Phys. Rev. B {\bf 77}, 125327 (2008).

\bibitem{Vojta} M. Vojta, R. Bulla, and W. Hofstetter, Phys. Rev. B {\bf 65}, 140405(R) (2002).

\bibitem{Zheng2002} B. Zheng, C.G. Lu, G. Gu, A. Makarovski, G. Finkelstein and J. Liu, Nano Lett. {\bf 2}, 895 (2002).

\bibitem{Kondo1998} D. Goldhaber-Gordon, H. Shtrikman, D. Mahalu, D. Abusch-Magder, U. Meirav and M.A. Kastner, Nature {\bf 391}, 156 (1998); S.M. Cronenwett, T.H. Oosterkamp and L.P. Kouwenhoven, Science {\bf 281}, 540 (1998); J. Schmid, J. Weis, K. Eberl and K. Von Klitzing, Physica B {\bf 258}, 182 (1998).

\bibitem{QDreview} L.P. Kouwenhoven, C.M. Marcus, P.L. McEuen, S. Tarucha, R.M. Westervelt and N.S. Wingreen, in \textit{Mesoscopic Electron Transport}, ed. by L.P. Kouwenhoven, G. Schon, and L.L. Sohn (Kluwer, Dordrecht, 1997), pp. 105-214.

\bibitem{4peaks} W. Liang, M. Bockrath, and H. Park, Phys. Rev. Lett. {\bf 88}, 126801 (2002); M.R. Buitelaar, A. Bachtold, T. Nussbaumer, M. Iqbal, and C. Schönenberger, Phys. Rev. Lett. {\bf 88}, 156801 (2002).

\bibitem{Pustilnik&Glazman2004} L. I. Glazman and M. Pustilnik, in \textit{Nanophysics: Coherence and Transport}, ed, by H. Bouchiat, Y. Gefen, S. Gueron, G. Montambaux, and J. Dalibard (Elsevier, New York, 2005),  pp. 427-478.

\bibitem{dissipation} Yu. Bomze, H. Mebrahtu, I. Borzenets, A. Makarovski, and G. Finkelstein,  Phys. Rev. B {\bf 79}, 241402(R) (2009).

\bibitem{Zumbuhl} D.M. Zumbuhl, C.M. Marcus, M.P. Hanson and A.C. Gossard, Phys. Rev. Lett. {\bf 93}, 256801 (2004)

\bibitem{BabicGap} B. Babic, T. Kontos, and C. Schönenberger, Phys. Rev. B {\bf 70}, 235419 (2004).

\bibitem{Nygard} J. Nygard, W.F. Koehl, N. Mason, L. DiCarlo, and C.M. Marcus, arXiv:cond-mat/0410467 (2004).

\bibitem{Fuhrer} M. S. Fuhrer, B. M. Kim, T. Dürkop, and T. Brintlinger, Nano Letters {\bf 2}, 755 (2002).

\bibitem{2stage} P.S. Cornaglia and D.R. Grempel, Phys. Rev. B {\bf 71}, 075305 (2005); C.-H. Chung, G. Zarand, and P. Wolfle, Phys. Rev. B {\bf 77}, 035120 (2008).

\bibitem{note} 
The low bias conductance in this sample is affected by a zero-bias anomaly (ZBA) that originates in the resistive leads, which serve as built-in filters \cite{dissipation}. To factor it out, we normalize the data in Figs.~3(d),(e) by the ZBA \emph{vs.} $T$ dependence measured simultaneously in the neighboring even-electron valley [visible on the left side of Fig.~3(e)], which is not affected by the Kondo box effect. The validity of the procedure is supported by the fact that the conductance in the even valley becomes independent of temperature for a range of $V_{\rm gate}$, as expected for temperatures much smaller than the relevant energy scales. 

\bibitem{2stage_expt} W.G. van der Wiel, S. De Franceschi, J.M. Elzerman, S. Tarucha, L.P. Kouwenhoven, J. Motohisa, F. Nakajima, and T. Fukui, Phys. Rev. Lett. {\bf 88}, 126803 (2002); G. Granger, M.A. Kastner, I. Radu, M.P. Hanson, and A.C. Gossard, Phys. Rev. B {\bf 72}, 165309 (2005).


\bibitem{Heersche2006} H.B. Heersche, Z. de Groot, J.A. Folk, L.P. Kouwenhoven, H.S.J. van der Zant,  A.A. Houck, J. Labaziewicz and I.L. Chuang, Phys. Rev. Lett. {\bf 96}, 017205 (2006).


\end{thebibliography}
\end{document}